\documentclass[prd,amssymb,aps,twocolumn,epsfig]{revtex4}
\usepackage{graphicx}% Include figure files
\usepackage{dcolumn}   % needed for some tables
\usepackage{amsmath,amssymb}
\usepackage{bm}

\def\kc#1{\left(#1\right)}

\def\be{\begin{equation}}       \def\ee{\end{equation}}
\def\bea{\begin{eqnarray}}      \def\eea{\end{eqnarray}}
\def\ba{\begin{array} }
\def\ea{\end{array} }
\def\bc{\begin{center}}
\def\ec{\end{center}}
\def\bnum{\begin{enumerate} }
\def\enum{\end{enumerate}}

\def\=>{\Rightarrow}
\def\>{\rightarrow}

\date{\today}

\begin{document}
%\draft
%\twocolumn[]
%\hsize
%\textwidth\columnwidth\hsize\csname@twocolumnfalse%
%\endcsname

\title{Anomalous gauge couplings of the Higgs boson at the CERN LHC: Semileptonic mode in WW scatterings}

%LHC Phenomenology of the effective Electroweak chiral field
%theory}

\author{Yong-Hui Qi$^{1,2}$, Yu-Ping Kuang$^{1,2}$, Bei-Jiang Liu$^{3}$, and Bin
Zhang$^{1,2}$}

\affiliation{$^1$ Department of Physics, Tsinghua University,
Beijing, 100084, China}

\affiliation{$^2$ Center for High Energy Physics, Tsinghua
University, Beijing, 100084, China}

\affiliation{$^3$ Institute of High Energy Physics, Academia Sinica, Beijing, 100039, China}
\begin{abstract}

We make a full tree level study of the signatures of anomalous gauge
couplings of the Higgs boson at the CERN LHC via the semileptonic
decay mode in WW scatterings, $pp\to W^+W^\pm j^f_1j^f_2\to
l^+\nu^{}_lj^{}_1j^{}_2j^f_1j^f_2$. Both signals and backgrounds are
studied at the hadron level for the Higgs mass in the range
$115~{\rm GeV}\le m^{}_H\le 200~{\rm GeV}$. We carefully impose
suitable kinematical cuts for suppressing the backgrounds. To the
same sensitivity as in the pure leptonic mode $pp\to W^+W^+
j^f_1j^f_2\to l^+\nu^{}_ll^+\nu^{}_lj^f_1j^f_2$, our result shows
that the semileptonic mode can reduce the required integrated
luminosity by a factor of 3. If the anomalous couplings in nature
are actually larger than the sensitivity bounds shown in the text,
the experiment can start the test for an integrated
luminosity of 50 fb$^{-1}$.\\\\
%they may even be detected for an integrated luminosity of 50 fb$^{-1}$.\\\\
\null\noindent{PACS numbers: 12.60.Fr, 11.15.Ex, 14.80.Cp}
\end{abstract}

\null\noindent{\null\hspace{5cm}TUHEP-TH-08166}

\maketitle

\section{Introduction}

Although the standard model (SM) has passed all the LEP electroweak
precision tests, its spontaneous symmetry breaking sector is still a
puzzle. The Higgs boson has not been found yet. The LEP direct
search bound on the SM Higgs mass is $m^{}_H>114.4$ GeV \cite{PDG},
and the $95\%$ CL upper bound on $m^{}_H$ from the LEP precision
data is $m^{}_H\le 167$ GeV \cite{PDG}. This range of the SM Higgs
mass is within the coverage of the CERN Large Hadron Collider (LHC),
and searching for the Higgs boson is of first priority in LHC
experiments. Theoretically, the SM Higgs sector suffers from the
well-known problems of {\it triviality} \cite{triviality} and {\it
unnaturalness} \cite{unnaturalness}. Therefore there must be a scale
of new physics, $\Lambda$, above which the SM should be replaced by
certain new physics model. Naturalness implies that $\Lambda\sim
O{\rm (TeV)}$. Direct search for the new heavy particle(s) with mass
$M\ge\Lambda$ at the LHC may or may not be easy depending on how
high $\Lambda$ actually is and their properties.
% of the new particle(s).
However, they will affect the couplings between lighter
particles through virtual processes. Once a light Higgs boson
candidate is found at the LHC, the first question to be answered is
whether it is the SM Higgs boson or a Higgs boson in certain new
physics model. The contribution of new heavy particles to the
couplings related to the Higgs boson will cause the couplings
anomalous (different from the SM values), therefore measuring the
anomalous Higgs couplings can answer the above question. The
anomalous couplings of the Higgs boson to electroweak (EW) gauge
bosons are of special interest since they are related to the mass
generation mechanism of the $W$ and $Z$ bosons. In this paper, we
concentrate on studying sensitive processes for measuring those
anomalous coupling constants at the LHC.

Since we do not know what the new physics model above $\Lambda$
really is, we study it in a general model independent way. There
have been various formulations describing the effective anomalous
couplings between the Higgs boson and the EW gauge bosons, namely
the linear realization formulation \cite{Hagiwara,Buchmuller,G-G}
and the nonlinear realization formulation \cite{C-K}. In this paper,
we take the popular linear realization formulation given in
\cite{Hagiwara,G-G} to perform the study. In this formulation, the
main anomalous gauge couplings of the Higgs boson deviating from the
SM coupling are of dimension six. The $CP$ conserving effective
Lagrangian for the anomalous interactions is formulated as
\cite{Hagiwara,G-G}
\begin{equation}                    %(1)
{\cal L}_{eff} ~\,=~\, \sum_n \frac{f_n}{\Lambda^2} {\cal O}_n \,,
\label{l:eff}
\end{equation}
where $f_n$'s are dimensionless {\emph{anomalous couplings}}. In the
SM, $f_n=0$. The gauge-invariant dimension six operators ${\cal
O}_n$'s are \cite{G-G}
\begin{eqnarray}                    %(2)
&&{\cal O}_{BW} =  \Phi^{\dagger} \hat{B}_{\mu \nu}
\hat{W}^{\mu \nu} \Phi, \nonumber \\
&& {\cal O}_{DW} = \mbox{Tr}([D_{\mu},\hat{W}_{\nu\rho}],[D^{\mu},\hat{W}^{\nu\rho}]), \nonumber \\
&&{\cal O}_{DB}=-\frac{{g^\prime}^2}{2} (\partial_\mu
B_{\nu\rho}) (\partial^\mu B^{\nu\rho}),\nonumber \\
&&{\cal O}_{\Phi,1} =  (D_\mu \Phi)^\dagger \Phi^\dagger \Phi
(D^\mu \Phi), \nonumber \\
&&{\cal O}_{\Phi,2} =\frac{1}{2}
\partial^\mu\kc{\Phi^\dagger \Phi}
\partial_\mu\kc{\Phi^\dagger \Phi}, \nonumber \\
&&{\cal O}_{\Phi,3} =\frac{1}{3} (\Phi^\dagger \Phi)^3,\nonumber\\
&& {\cal O}_{WWW}=\mbox{Tr}[\hat{W}_{\mu
\nu}\hat{W}^{\nu\rho}\hat{W}_{\rho}^{\mu}]
, \nonumber \\
&& {\cal O}_{WW} = \Phi^{\dagger} \hat{W}_{\mu \nu}
\hat{W}^{\mu \nu} \Phi , \nonumber \\
&&{\cal O}_{BB} = \Phi^{\dagger} \hat{B}_{\mu \nu} \hat{B}^{\mu
\nu} \Phi ,  \nonumber \\
&&{\cal O}_W  = (D_{\mu} \Phi)^{\dagger}
\hat{W}^{\mu \nu}  (D_{\nu} \Phi), \nonumber \\
&&{\cal O}_B  =  (D_{\mu} \Phi)^{\dagger} \hat{B}^{\mu \nu}
(D_{\nu} \Phi),
\label{O}
\end{eqnarray}                          %(3)
where $\hat B_{\mu\nu}$ and $\hat W_{\mu\nu}$ stand for
\begin{eqnarray}
\hat{B}_{\mu \nu} = i \frac{g'}{2} B_{\mu \nu},\;\;\;\;\;\;\;\;
\hat{W}_{\mu \nu} = i \frac{g}{2} \sigma^a W^a_{\mu \nu},
\label{B,W}
\end{eqnarray}
in which $g$ and $g^\prime$ are the $SU(2)$ and $U(1)$ gauge
coupling constants, respectively.

It has been shown that the operators $O_{\Phi,1}$, $O_{BW}$,
$O_{DW}$, $O_{DB}$ are related to the two-point functions of the
weak bosons, so that they are severely constrained by the precision
EW data \cite{G-G}. For example, $O_{BW}$ and $O_{\Phi,1}$ are
related to the oblique correction parameter $S$ and $T$, and are
thus strongly constrained by the precision EW data. The $2\sigma$
constraints on $|f_{BW}/\Lambda^2|$ and $|f_{\Phi,1}/\Lambda^2|$
are: $|f_{BW}/\Lambda^2|, |f_{\Phi,1}/\Lambda^2|<
O(10^{-2})$~TeV$^{-2}$ \cite{ZKHY03}. The operators $O_{\Phi,2}$ and
$O_{\Phi,3}$ are related to the triple and quartic Higgs boson
self-interactions, and have been studied in detail in
Ref.~\cite{BHLMZ}. The operator $O_{WWW}$ is related to the weak
boson self-couplings, so that it is irrelevant to the present study.
The precision and low energy EW data are not sensitive to the
remaining four operators $O_{WW}$, $O_{BB}$, $O_{W}$, and $O_{B}$.
These four anomalous couplings are only constrained by the
requirement of the unitarity of the $S$ matrix, and such theoretical
constraints are quite weak \cite{unitarity}. For example, the
unitarity constraints on $f_W/\Lambda^2$ and $f_{WW}/\Lambda^2$ are
\cite{unitarity,ZKHY03}:
\begin{eqnarray}                            %(4)
\left|\frac{f_W}{\Lambda^2}\right|\le 7.8~{\rm TeV}^{-2}, ~~~~~~
\left|\frac{f_{WW}}{\Lambda^2}\right|\le 39.2~{\rm TeV}^{-2}.
\label{unitarity bounds}
\end{eqnarray}
The test of these four anomalous Higgs couplings at the LHC is what
we shall concentrate on. The sensitivity of the test is crucial for
discriminating models.

Taking account of the mixing in the neutral gauge boson sector, the effective Lagrangian
expressed in terms of the photon field $A_\mu$, the weak boson fields $W^\pm_\mu$, $Z_\mu$,
and the Higgs boson field $H$ is \cite{G-G}
\begin{eqnarray}                              %(5)
&&\hspace{-0.4cm}{\cal L}^H_{\rm eff}=g^{}_{H\gamma\gamma}HA_{\mu\nu}A^{\mu\nu}
+g^{(1)}_{HZ\gamma}A_{\mu\nu}Z^\mu\partial^\nu H\nonumber\\
&&%\hspace{0.4cm}
+g^{(2)}_{HZ\gamma}HA_{\mu\nu}Z^{\mu\nu}
+g^{(1)}_{HZZ}Z_{\mu\nu}Z^\mu\partial^\nu H\nonumber\\
&&%\hspace{0.4cm}
+g^{(2)}_{HZZ}HZ_{\mu\nu}Z^{\mu\nu}
+g^{(1)}_{HWW}(W^+_{\mu\nu} W^{-\mu}\partial^\nu H+{\rm h.c.})\nonumber\\
&&%\hspace{0.4cm}
+g^{(2)}_{HWW}HW^+_{\mu\nu}W^{-\mu\nu},
\label{LHeff}
\end{eqnarray}
where the anomalous couplings $g^{(i)}_{HVV}$ with~$i=1,2$ ($V_\mu$
stand for $A_\mu,~W^\pm_\mu$ or $Z_\mu$) are related to the
anomalous couplings $f_n$'s by
\begin{eqnarray}                           %(6)
&&\hspace{-0.4cm}g^{}_{H\gamma\gamma}=-\kappa\frac{s^2(f_{BB}
+f_{WW})}{2},\nonumber\\
&&\hspace{-0.4cm}g^{(1)}_{HZ\gamma}=\kappa\frac{s(f_W-f_B)}{2c},
%\nonumber\\
%&&
~~~~g^{(2)}_{HZ\gamma}=\kappa\frac{s[s^2f_{BB}
-c^2f_{WW}]}{c},\nonumber\\
&&\hspace{-0.4cm}g^{(1)}_{HZZ}=\kappa\frac{c^2f_W+s^2f_B}{2c^2},
%\nonumber\\
%&&
~~g^{(2)}_{HZZ}=-\kappa\frac{s^4f_{BB}
+c^4f_{WW}}{2c^2},\nonumber\\
&&\hspace{-0.4cm}g^{(1)}_{HWW}=\kappa\frac{f_W}{2},%\nonumber\\
%&&
~~~~~~~~~~~~~g^{(2)}_{HWW}=-\kappa{f_{WW}}, \label{g}
\end{eqnarray}
in which $s\equiv \sin\theta_W,~c\equiv \cos\theta_W$ and
$\displaystyle\kappa\equiv\frac{gM^{}_W}{\Lambda^2}\approx{0.053}\left(\frac{1~{\rm
TeV}}{\Lambda}\right)^2$ TeV$^{-1}$.

Once nonvanishing values of these anomalous couplings (after
subtracting the corresponding SM loop corrections) are detected
experimentally, it implies that we have already seen the effect of
new physics beyond the SM. There have been papers studying the test
of the above four anomalous Higgs couplings at the LHC
\cite{LHC,Zeppenfeld,ZKHY03}, the linear collider \cite{LC,BHLMZ},
and the photon colliders \cite{HKZ06}. So far the most sensitive
test at the LHC is via the pure leptonic mode in $W^+W^+$
scattering, $pp\to W^+W^+ j^f_1j^f_2\to
l^+\nu^{}_ll^+\nu^{}_lj^f_1j^f_2$ ($j^f_1j^f_2$ are the two forward
jets characterizing $WW$ fusion). This process is sensitive in
testing the anomalous couplings $f_W$ and $f_{WW}$ but less
sensitive in testing $f_B$ and $f_{BB}$ \cite{ZKHY03}. The obtained
$3\sigma$ constraints for an integrated luminosity of 300 fb$^{-1}$
on $f_W$ and $f_{WW}$ are \cite{ZKHY03}:
\begin{eqnarray}                               %(7)
\frac{|f_{W}|}{\Lambda^2}\le 1.6~ {\rm TeV}^{-2}, ~~~~~~
\frac{|f_{WW}|}{\Lambda^2}\le 2.9~ {\rm TeV}^{-2}.
\label{LHC:3sigma}
\end{eqnarray}
We see that these values are significantly smaller than the unitarity
bounds (\ref{unitarity bounds}), so that there is plenty of room for detectable
$f_W/\Lambda^2$ and $f_{WW}/\Lambda^2$ within the unitarity bounds.

However, the required integrated luminosity 300 fb$^{-1}$ is rather
high. The LHC needs several years after its first collision to reach
this high integrated luminosity. In this paper, we study the
possibility of taking the semileptonic mode which can have a larger
cross section. Since it is not possible to distinguish $W^+\to
j_1j_2$ and $W^-\to j_1j_2$ experimentally, we have to study the
scatterings $pp\to W^+W^\pm j^f_1j^f_2$ with $W^+\to
l^+\nu^{}_l,~W^\pm\to j_1j_2$. There are four jets in the final
state, so that the study of the signal and backgrounds is much more
complicated than that in the pure leptonic mode. We have to
calculate at the hadron level rather than the parton level. We shall
show that, from a detailed study, certain kinematic cuts can
suppress the backgrounds, and the required integrated luminosity for
reaching the $3\sigma$ sensitivity (\ref{LHC:3sigma}) can be reduced
to 100 fb$^{-1}$. If the anomalous couplings in the real world are
not so small, say larger than the $1\sigma$ bounds $-3.5~{\rm
TeV}^{-2}\le f^{}_W/\Lambda^2\le 1.3~{\rm TeV}^{-2}$ or $-0.9~{\rm
TeV}^{-2}\le f^{}_{WW}/\Lambda^2\le 0.8~{\rm TeV}^{-2}$ , the LHC
can already detect their effect when the integrated luminosity
reaches 50 fb$^{-1}$. If they are larger than the bounds $-4.5~{\rm
TeV}^{-2}\le f^{}_W/\Lambda^2\le 2.4$ TeV$^{-2}$ or $-2.0~{\rm
TeV}^{-2}\le f^{}_{WW}/\Lambda^2\le 1.5$ TeV$^{-2}$, a 3$\sigma$
detection can be performed at the LHC for an integrated luminosity
of 50 fb$^{-1}$.

This paper is organized as follows. In Sec. II, we briefly sketch
some key points in the calculation of weak boson scatterings at the
LHC. All the main backgrounds and kinematic cuts for suppressing the
backgrounds are investigated in Sec. III. The numerical results of
the cross sections and detecting sensitivities under the imposed
kinematic cuts are presented in Sec. IV. Sec. V is a concluding
remark.

\section{Weak Boson Scatterings}

Weak boson scatterings ($VV\to VV$) at the LHC are usually regarded
as useful processes for probing strongly interacting electroweak
symmetry breaking (EWSB) mechanism, and have been studied in details
\cite{Bagger9495}. In addition, even if EWSB is driven by light
Higgs boson, it has been shown that $VV\to VV$ also provide
sensitive tests of the anomalous gauge couplings of the Higgs boson
\cite{ZKHY03}. Some anomalous gauge couplings of the Higgs boson may
be first detected in on-shell Higgs productions to a lower
sensitivity \cite{Zeppenfeld}. Weak boson scatterings can then
provide further sensitive tests to get more useful information about
new physics.

\begin{figure}[t]                           %Fig.1
%\centering \hspace*{\fill}\bc
%\vspace{-0.5cm}
\includegraphics[scale=0.75]{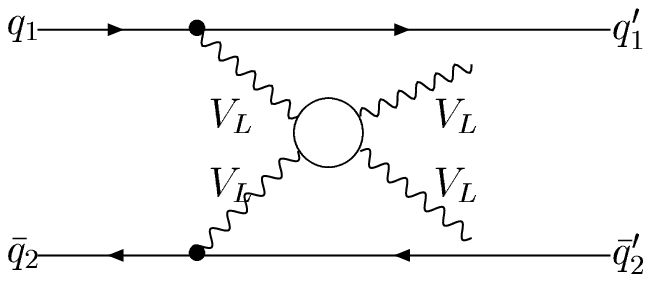}\\%{VVscatteringTot.eps}
\hspace{0.3cm}(a)\\
\vspace{0.4cm}\includegraphics[scale=0.65]{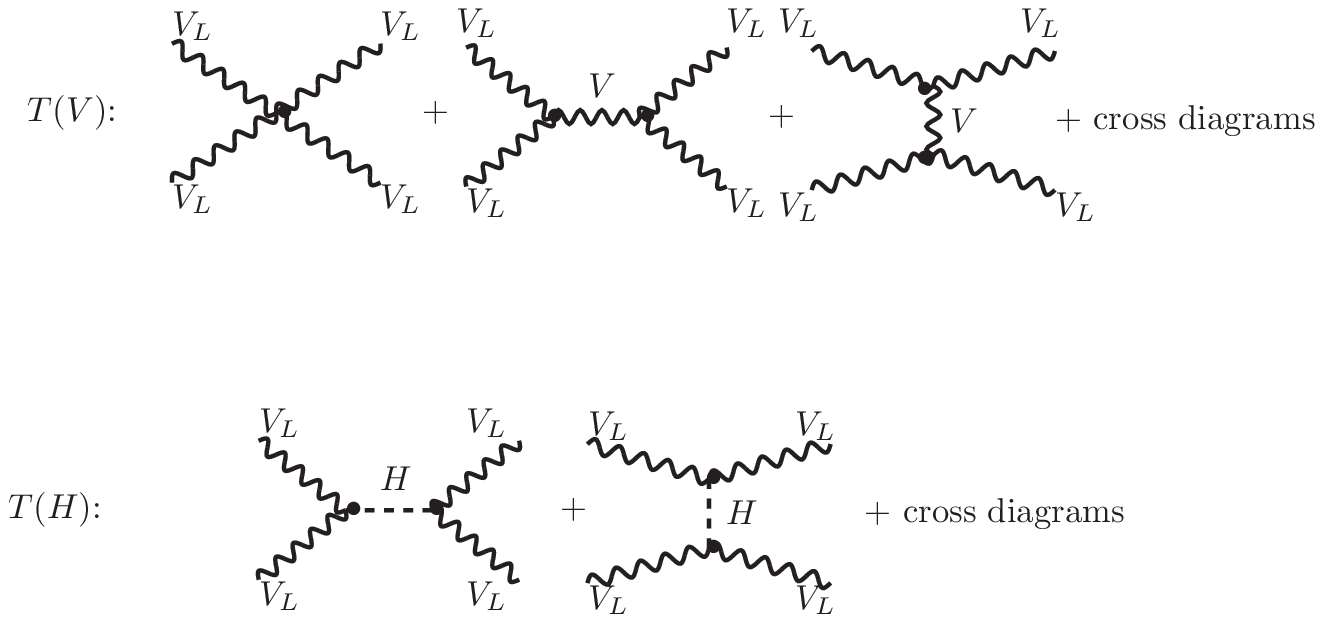}%{VVscattering.eps}
\hspace{0.5cm}(b) \null\vspace{-0.3cm} \caption{ (a) symbolic
diagrams for weak boson scatterings. (b) the two kinds of scattering
amplitudes $T(V)$ and $T(H)$ in weak boson scatterings.}
\label{VV-VV}
\end{figure}

In weak boson  scatterings (cf. FIG.\,\ref{VV-VV}(a)), a quark $q_1$
in a proton becomes a forward jet $j^f_1$ (from the outgoing quark
$q'_1$) after emitting a weak boson. It can be seen from helicity
analysis that, if $j^f_1$ and $j^f_2$  are sufficiently forward, the
emitted weak bosons are mainly longitudinal. So that the "initial
state" weak bosons in FIG.\,\ref{VV-VV} are $V_L$'s. Let us look at
the longitudinal weak boson scatterings $V_LV_L\to V_LV_L$. At tree
level, there are two kinds of weak boson scattering amplitudes shown
in FIG.\,\ref{VV-VV}(b), namely the amplitude containing only gauge
bosons $T(V)$, and the amplitude containing Higgs boson exchanges
$T(H)$. Since the longitudinal polarization vector depends on the
momentum of $V_L$, the two amplitudes $T(V)$ and $T(H)$ all depend
on the center of mass energy $E$ as $E^2$. In the SM, the coupling
constant between the Higgs boson and weak bosons in $T(H)$ is the
same as the gauge coupling constant $g$ in $T(V)$. This makes the
$E^2$-dependence terms in $T(V)$ and $T(H)$ exactly cancel in the
total amplitude $T(V)+T(H)$, leading to a $E^0$-dependence of the
total amplitude, which guarantees the unitarity of the $S$ matrix.
In the case that the $HVV$ couplings in $T(H)$ are anomalous, the
cancellation will not be exact, which leads to a $E^2$-dependence of
the total amplitude. The magnitude of the remained $E^2$-dependence
depends on the size of the anomalous couplings. So far as the
anomalous couplings are within the unitarity bounds (\ref{unitarity
bounds}), there is no violation of the unitarity of the $S$ matrix
below the new physics scale $\Lambda$. Thus in the high energy
region of the LHC, the cross section is quite different from that in
the SM. This is the reason why weak boson scatterings provide
sensitive tests of the anomalous couplings. Different from the case
of testing the strongly interacting EWSB mechanism in
Ref.~\cite{Bagger9495}, the signal in the present case is defined as
the cross section with anomalous couplings $f_n\ne0$ rather than the
longitudinal cross section. So the $V_LV_L\to V_LV_T,V_TV_T$
contributions with $f_n\ne0$ are also signals. However, the
transverse polarization vector is not momentum dependent, so that
the $V_LV_L\to V_LV_L$ contribution with $f_n\ne0$ is the most
sensitive signal.

At the parton level, the signals and backgrounds in the gold-plated
pure leptonic modes of weak boson scatterings have been studied
systematically in Ref.~\cite{Bagger9495}. Studying at the parton
level, Ref.~\cite{ZKHY03} showed that the $W^+_LW^+_L\to W^+_LW^+_L$
process is the most sensitive one for testing the anomalous
couplings (\ref{g}). Now we are going to study the semileptonic mode
with $W^+W^+\to l^+\nu^{}_lj_1j_2$. Since it is not possible to
distinguish the jets from $W^+_L\to j_1j_2$ and $W^-_L\to j_1j_2$
experimentally, we have to take account of both the $W^+_LW^+_L$ and
$W^+_LW^-_L$ productions and tag the final state $W^+_LW^\pm_L\to
l^+\nu^{}_lj_1j_2$. So we are going to calculate the full tree level
contributions to the process
\begin{eqnarray}                              %(8)
pp\to W^+W^\pm j^f_1j^f_2\to l^+\nu^{}_lj_1j_2j^f_1j^f_2,
\label{full process}
\end{eqnarray}
where $W^+$ and $W^\pm$ are on-shell. Now the final state contains
four jets, namely the two forward jets $j^f_1j^f_2$ and the two jets
$j_1j_2$ from $W^\pm$ decays, so that the parton level study is not
sufficient for finding out the suitable kinematic cuts to suppress
the large backgrounds.

In the following, we shall work at the hadron level, calculating the
full tree level contributions to the signal and backgrounds using
the helicity amplitude methods \cite{helicity} and the package
PYTHIA \cite{PYTHIA} with its default fragmentation model. For the
parton distribution functions, we take CTEQ6L \cite{CTEQ6L}. For the
reconstruction of the $W$ boson from the two jets $j_1j_2$, we take
the cluster-type jet algorithm \cite{Catani}, and using the package
ALPGEN \cite{ALPGEN}. We shall develop suitable kinematic cuts to
suppress the backgrounds and save the signal as much as possible.

The backgrounds to $V_LV_L$ scatterings can be classified into three kinds, namely the
EW background, the QCD background, and the top quark background \cite{Bagger9495}. The
irreducible EW background amplitudes (with the same final state particles as the signal)
should be calculated together with the signal amplitude to
guarantee gauge invariance. %The related Feynman diagrams are illustrated in FIG.~\ref{signal-EWbkgd}.
Other backgrounds with different initial or final state particles
can be calculated separately.

Let $\sigma(f_n\ne 0)$ and $\sigma^{}_B\equiv \sigma(f_n=0)$ be the
total and background cross sections, respectively. We define the
signal cross section $\sigma^{}_S$ by
\begin{eqnarray}                               %(9)
\sigma^{}_S\equiv\sigma(f_n\ne 0)-\sigma^{}_B.
\label{sigma_S}
\end{eqnarray}

Now the main experimental interest is to find out new physics effect
beyond the SM background. Let $N^{}_{S}$ and $N^{}_B$ be the numbers
of the signal events and background events, respectively. For large
values of $N^{}_S$ and $N^{}_B$, we determine the statistical
significance $\sigma^{}_{stat}$ according to
\begin{eqnarray}                                 %(10)
\sigma^{}_{stat}=\frac{N^{}_S}{\sqrt{N^{}_B}}.
\label{significance}
\end{eqnarray}

However, the simple expression (\ref{significance}) holds only when
$N^{}_S$ and $N^{}_B$ are large. For general values of $N^{}_S$ and
$N^{}_B$, (\ref{significance}) is not precise enough, and we should
take the general Poisson probability distribution approach
\begin{eqnarray}                           %(11)
\displaystyle
P^{}_B&=&\sum_N\displaystyle e^{-N^{}_B}\frac{N_B^N}{N!},\nonumber\\
&&N=N^{}_S+N^{}_B,N^{}_S+N^{}_B+1,\cdots,\infty.
\label{P_B}
\end{eqnarray}
From the obtained value of $1-P^{}_B$, we can find out the
corresponding value of $\sigma^{}_{stat}$ \cite{PDG}. The value of
$\sigma^{}_{stat}$ obtained in this way approaches to that given in
(\ref{significance}) when $N^{}_S$ and $N^{}_B$ are sufficiently
large. We shall take the approach (\ref{P_B}) throughout this paper.

\begin{center}
\begin{figure}[b]                                  %Fig.2
\includegraphics[width=8.4truecm,clip=true]{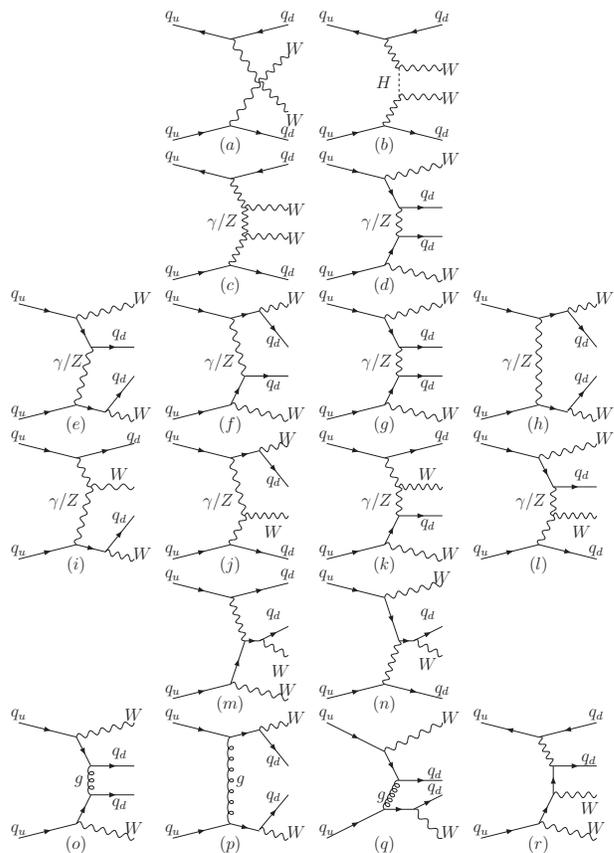}
\vspace{-0.5cm} \caption{Typical Feynman diagrams for the signal and
irreducible backgrounds in $W^+W^+$ scattering.}
\label{signal-EWbkgd}
\end{figure}
\end{center}

\null\vspace{-1.2cm}

\section{Backgrounds and Cuts}

Now we consider all the three kinds of backgrounds to $pp\to
W^+W^\pm j^f_1j^f_2\to l^+\nu_lj_1j_2j^f_1j^f_2$, and study suitable
kinematic cuts for suppressing them.

Considering the actual acceptance of the detectors at the LHC, we
always require all the final state particles to be in the following
rapidity range throughout this paper
\begin{eqnarray}                               %(12)
|\eta|<4.5.
\label{acceptance}
\end{eqnarray}

Recently, Ref.~\cite{Butterworth} provided a systematic hadron level
study of the semileptonic modes in $WW$ scatterings at the LHC for
testing the EW chiral Lagrangian coefficients when there are heavy
resonances enhancing the scattering cross section at high energies.
Although we assume there is no heavy resonances in our present case,
the cross section is also enhanced at high energies by the energy
dependence arising from the anomalous couplings. Thus the new
techniques developed in Ref.~\cite{Butterworth} are also useful in
our case. We shall apply some of their techniques to our study of
testing the anomalous couplings of the light Higgs boson.

\subsection{Signal and Irreducible Backgrounds}

As mentioned above that the signal and irreducible background
amplitudes should be put together in the calculation to guarantee
gauge invariance. Take the $pp\to W^+W^+ J^f_1j^f_2$ process as an
example. The typical Feynman diagrams for these amplitudes are shown
in FIG.\,\ref{signal-EWbkgd} in which FIG.\,\ref{signal-EWbkgd}(b)
(containing Higgs boson exchange) is the signal, and the total
contribution of these diagrams with $f^{}_n=0$ is the irreducible
backgrounds.

The final state particles in the signal process contains two forward jets $j^f_1j^f_2$, two
jets $j_1j_2$ from $W^\pm$ decays, a positively charged lepton $l^+$ and a missing neutrino $\nu^{}_l$.
Let us consider the cuts for each of the final state particles for extracting the signal.

\subsubsection{Charged Lepton and Forward Jets}

\null\vspace{-0.9cm}
\begin{figure}[h]                            %Fig.3
\bc \includegraphics[width=8.5cm,height=6cm]%scale=0.75]
{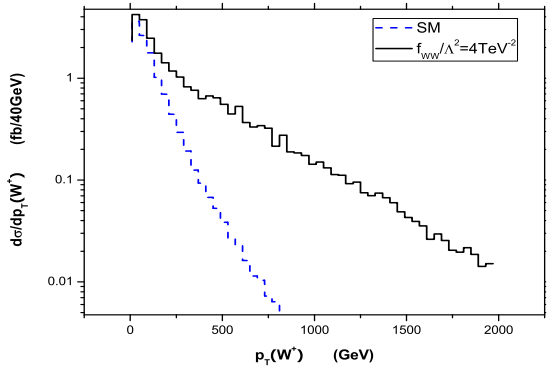}%\\
 \ec\null\vspace{-1.2cm}
 \caption{$d\sigma/dp^{}_T(W^+)$ distributions:
The solid and dashed curves stand for the cases of
$f_{WW}/\Lambda^2=4~{\rm TeV}^{-2},~f^{}_W/\Lambda^2=0$ and
$f^{}_{WW}/\Lambda^2=f^{}_W/\Lambda^2=0$, respectively.}
\label{p_T(W^+)}
\end{figure}

Let us first consider the cut for the transverse momentum of the
charged lepton $l^+$.  Since the $W^+$ boson is quite energetic, the
charged lepton $l^+$ moves almost along the direction of $W^+$. So
we can look at the transverse momentum distribution of $W^+$. Take
the case of $f^{}_{WW}$ dominant as an example.  FIG.\,\ref{p_T(W^+)}
shows the transverse momentum distributions of the $W^+$ decaying
into leptons with $f_{WW}/\Lambda^2=4~{\rm
TeV}^{-2},f^{}_W/\Lambda^2=0$ and with
$f^{}_{WW}/\Lambda^2=f^{}_W/\Lambda^2=0$ (the irreducible
background), respectively. We see that the distribution including
the signal is significantly harder than that of the irreducible
background. Thus we know that the transverse momentum distribution
of the signal $l^+$ is significantly harder than that of the
background $l^+$. From FIG.\,\ref{p_T(W^+)}, we see that imposing the
following $p^{}_T(l^+)$ cut can suppress the irreducible background
and keep the signal as much as possible,
\begin{eqnarray}                         %(13)
&&p^{}_T(l^+)>200~{\rm GeV} \label{p_T(l^+)}.
\end{eqnarray}

 After the cut (\ref{p_T(l^+)}), the jets in
most of the irreducible background processes are mainly in the low
$|\eta|$ region. Thus imposing the requirement of the forward jets
will effectively suppress this backgrounds. The observation of the
tagging forward jets do not depend on whether we are testing the
strongly interacting EWSB mechanism or testing the anomalous
couplings of a light Higgs boson. So we can follow
Ref.~\cite{Butterworth} to
%take the forward jet cuts. Following
%Ref.~\cite{Butterworth}, we
impose the following cuts on the
transverse momentum $p^{}_T(j^f)$, the energy $E(j^f)$, and the
pseudorapidity $\eta(j^f)$ of the two tagging forward jets
\cite{Butterworth}.
\begin{eqnarray}                                %(14)
&& p^{}_T(j^f_i)> 20~{\rm GeV},~~~~
E(j^f_i)>300~{\rm GeV},\nonumber\\
&&2.0<|\eta(j^f_i)|<4.5,~~i=1,2,\nonumber\\
&&\eta(j^f_1)\eta(j^f_2)<0. \label{forward jet cuts}
\end{eqnarray}
The rapidity cuts in (\ref{forward jet cuts}) guarantee the two
forward jets moving almost back-to-back. Later, we shall see that
this forward jet cut will also suppress the $W+jets$ QCD background
and the top quark background effectively. The efficiency of these
cuts are listed in the second and third rows in TABLE\,\ref{cut
efficiency}. We see that the cuts (\ref{p_T(l^+)}) and (\ref{forward
jet cuts}) can suppress the irreducible background quite
effectively.

\subsubsection{Hadronic Decay of the $W$ boson}

Now we come to the issue of extracting the $W^\pm\to j_1j_2$ events.
Since the final state $W^\pm$ is very energetic, $98\%$ of the two
jets $j_1j_2$ behave like a ``single'' energetic jet $J$ along the
$W^\pm$ direction \cite{Butterworth}, we first use the $k_T$
algorithm (the ALPGEN package \cite{ALPGEN}) with $E$ combination to
pick up the most energetic ``single jet''. Since $W^\pm$ and $W^+$
are almost back-to-back, we can impose the following cuts
\begin{eqnarray}                                     %(15)
 p^{}_T(J)>200~{\rm GeV},~~~~~~\eta(J)\eta(l^+)<0,
 \label{p_T(J)}
 \end{eqnarray}
 and requiring the invariant mass $M_J$ to reconstruct the $W^\pm$
mass, i.e.
\begin{eqnarray}                                      %(16)
 65~{\rm GeV}<M_J<95~{\rm GeV}, \label{M_J}
\end{eqnarray}

\null\noindent in which we have considered the realistic detection
resolution $\pm15$ GeV \cite{ATLAS}.

\subsection{QCD Backgrounds}

\begin{figure}[h]                                    %FIG.4
\centering \bc
\includegraphics[width=7.5cm,height=4.4cm]{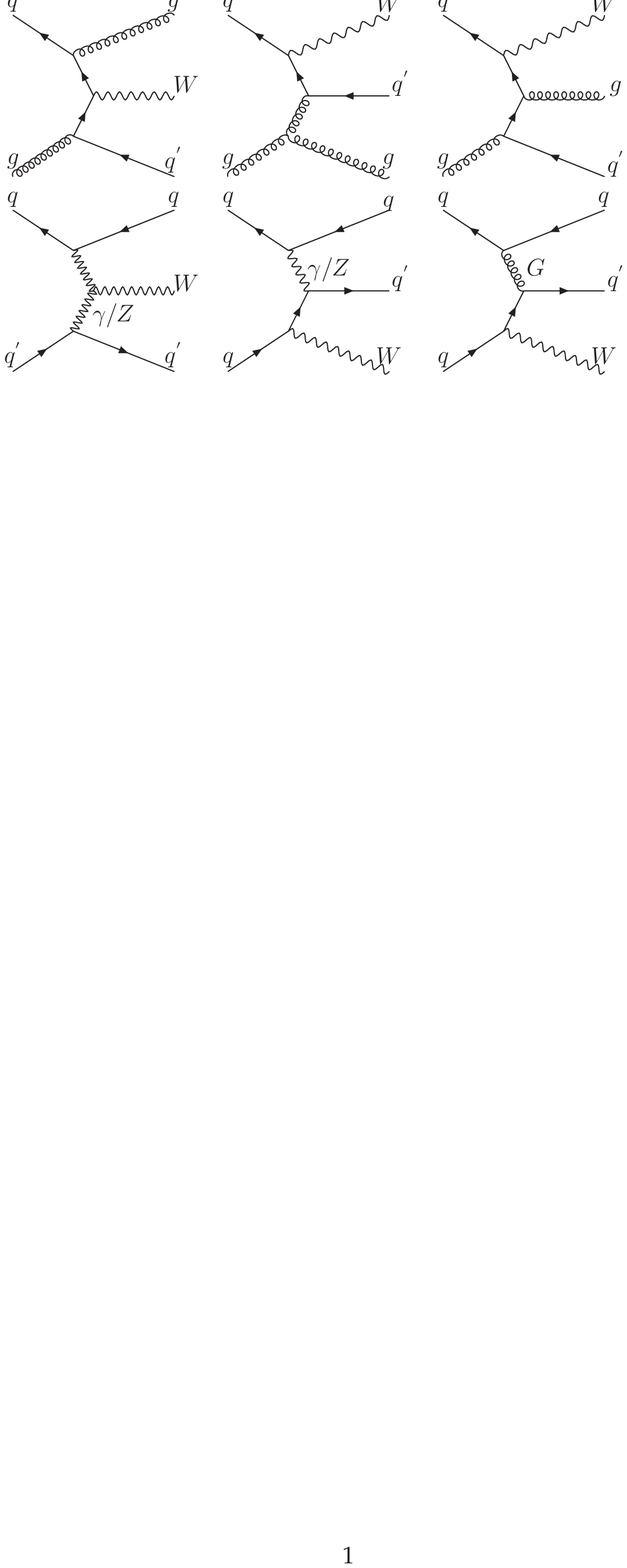}
\ec \vspace{-0.5cm}\caption{\label{W+2-parton} Typical diagrams for
$W+$ 2-parton.} %\label{W+2-parton}
\end{figure}

One of the important QCD backgrounds is $pp\to W+\hat n$-parton
which may leads to the final state $W+n$-jet at the hadron level.
The case that three of the $n$ jets are detected (with other jets
undetected), will be a background to the signal. We have examined
the cases for $\hat n=1,2,3,4$ and found that the most important
background comes from $\hat n=2$. Thus the main QCD background of
this kind is
\begin{eqnarray}                                  %(17)
pp\to W+2{\rm -parton}\to W+3{\rm -jet}. \label{W+3jets}
\end{eqnarray}
The typical Feynman diagrams for $qq,qg\to W+2{\rm-parton}$ are
depicted in FIG.\,\ref{W+2-parton}.

\null\vspace{-0.2cm}

Another similar QCD background is
\begin{eqnarray}                                  %(18)
%&&q\bar q,qg,gg\to W+n{\rm -}jet\nonumber\\
q\bar q,qg,gg\to WW+n{\rm -jet}. \label{WW+jets}
\end{eqnarray}

As mentioned above, the jets in the backgrounds (\ref{W+3jets}) and
(\ref{WW+jets}) are less forward than the forward jets in the signal
process when the lepton $l^+$ is constrained by (\ref{p_T(l^+)}). So
imposing the cuts (\ref{p_T(l^+)}) and (\ref{forward jet cuts}) can
suppress these two kinds of QCD backgrounds effectively.
Furthermore, the requirements (\ref{p_T(J)}) and (\ref{M_J}) can
significantly suppress this kind of background.

\begin{figure}[h]                              %Fig.5
\includegraphics[width=8.5cm,height=5cm]{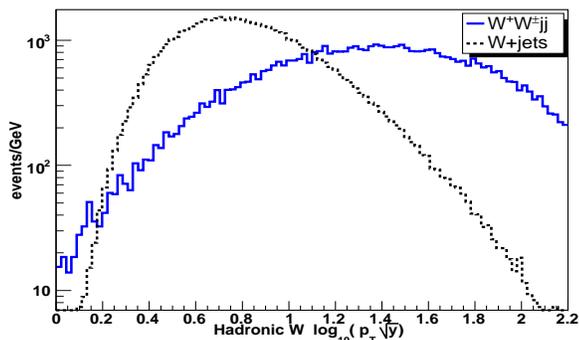}
\null\vspace{-0.4cm}\caption{ $\log(p^{}_T\sqrt{y})$ distributions
for the $pp\to W^+W^\pm j^f_1j^f_2$ and $pp\to W+3{\rm -jet}$
processes with $f_{W}/\Lambda^{2}=4$ TeV$^{-2}$,
$f^{}_W/\Lambda^2=0$. } \label{W-log}
\end{figure}

We can further impose a cut to suppress the above QCD backgrounds.
The $y$ cut method (imposing a cut on $\log (p^{}_T\sqrt{y})$)
developed in Ref.~\cite{Butterworth} is very effective for this
purpose. FIG.\,\ref{W-log} shows the $\log (p^{}_T\sqrt{y})$
distributions for the $pp\to W^+W^\pm j^f_1j^f_2$ (with
$f_W/\Lambda^2=4$ GeV$^{-2}$) and $pp\to W+3{\rm -jet}$
processes.
%\null\vspace{-0.8cm}
From FIG.\,\ref{W-log} we see that a cut \cite{Butterworth}
\begin{eqnarray}                                      %(19)
1.6<\log(p^{}_T\sqrt{y})<2.0 \label{log cut}
\end{eqnarray}
can effectively suppress the backgrounds. Indeed, after the cut
(\ref{p_T(J)}), (\ref{M_J}) and (\ref{log cut}), the above QCD
backgrounds are significantly reduced (cf. the fourth and fifth rows
in TABLE\,\ref{cut efficiency}).

\begin{figure}[h]                                   %FIG.6
\centering   \bc
\includegraphics[width=5.5cm,height=6.5cm]
{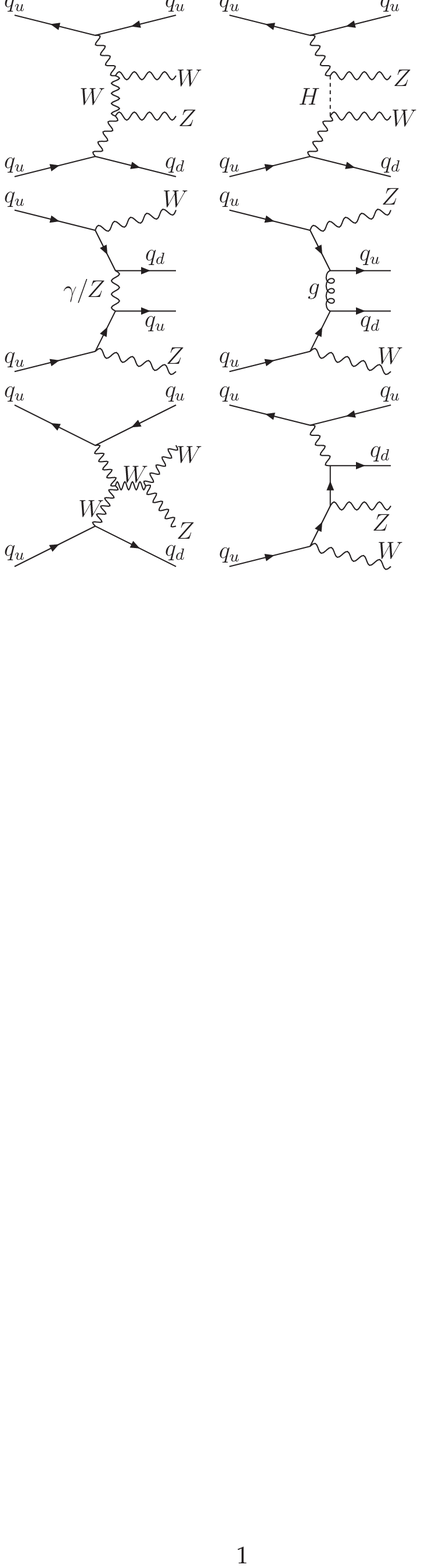} \ec\vspace{-0.7cm} \caption{ Typical diagrams
for the $WZ+2$-$jet$ background.} \label{WZ2j}
\end{figure}

 There is also a kind of important QCD background which is the
$WZjj$ process (cf. FIG.\,\ref{WZ2j}) since $M_Z$ is within the range
in (\ref{M_J}). This includes the $WZ$ scattering process, $pp\to
W^+Zj^f_1j^f_2$, which is quite similar to the signal process $pp\to
W^+W^\pm j^f_1j^f_2$. However, $M_Z$ is close to the upper bound in
(\ref{M_J}), i.e., a large portion of the tail of the $M_Z$
resonance higher than the peak is cut away by (\ref{M_J}), so that
the $WZ$ scattering background is significantly smaller than the
signal. However, there are processes of this kind other than $WZ$
scattering (cf. FIG.\,\ref{WZ2j}) which can be large. We see from the
fourth column of TABEL~\ref{cut efficiency} that all the cuts
imposed above can effectively suppress this kind of background.

%\null\vspace{-0.9cm}
\begin{figure}[h]                              %Fig.7
\includegraphics[width=8.5cm,height=5cm]%scale=0.32]
{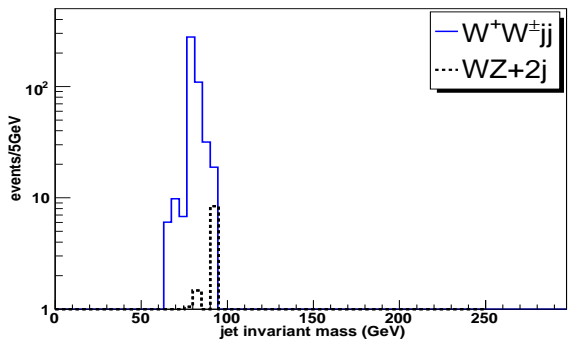}
\null\vspace{-0.4cm} \caption{Reconstruction of the $W^\pm$ boson
after the cuts (\ref{p_T(J)}) and (\ref{M_J}). The solid curve is
the $W^\pm$ peak in the signal process; the dashed curve shows the
$Z$ boson peak in the $WZ+2{\rm -jet}$ background.} \label{W
reconstruction}
\end{figure}

FIG.\,\ref{W reconstruction} shows the reconstructed $W$ boson peak
in the signal process and the $Z$ boson peak in the $WZ$ scattering
background after imposing the above cuts. We see that the $W$ boson
peak is clearly reconstructed, and the $Z$ boson peak is
significantly suppressed by the condition (\ref{M_J}).

\subsection{Top Quark Background}

\begin{figure}[h]                                 %Fig.8
\centering   \bc
\includegraphics[width=6.9cm,height=4.4cm]{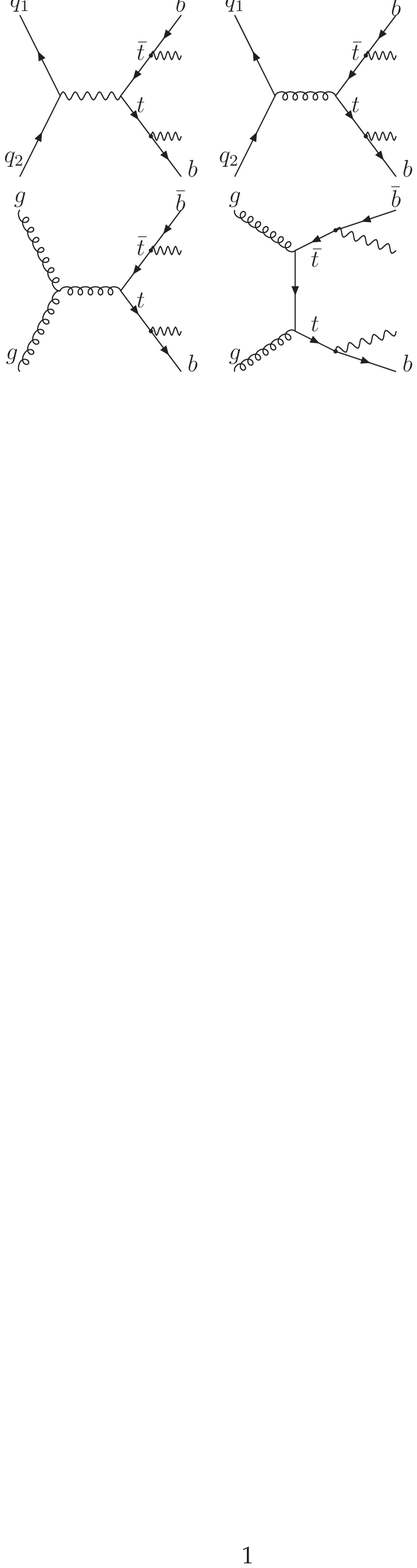}
\ec \vspace{-0.5cm}\caption{\label{ttbar-bkgd} Typical diagrams for
the $t\bar t$ background $gg\to t\bar t\to bW^+\bar{b}W^-$. }
\end{figure}

$W$ boson productions from the decay of top quarks in $t\bar t$
production (cf. FIG.\,\ref{ttbar-bkgd}) is an important background
which mimics the signal.
As mentioned above, the jets in this background are less forward
than the two forward jets in the signal, so that the forward jet
cuts (\ref{forward jet cuts}) can suppress this background.

%\null\vspace{-0.9cm}
\begin{figure}[h]                                 %Fig.9
\centering   \bc
\includegraphics[width=8.5cm,height=5.2cm]{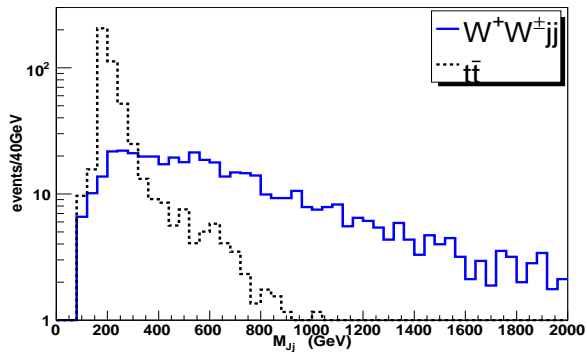}
\ec \vspace{-0.6cm}\caption{\label{M_{Jj}} Invariant mass $M_{Jj}$
distributions for the top quark background (dashed curve) and the
$pp\to W^+W^\pm j^f_1j^f_2$ process (solid curve). } \label{ttbar}
\end{figure}

%\null\vspace{-0.7cm}
However, further effective suppression is still needed. In the case
of pure leptonic mode, this can be significantly suppressed by
vetoing the central jets \cite{Bagger9495}. But in the semileptonic
mode, the signal $W^\pm\to j_1j_2$ is in the central rapidity
region, so that central jet veto cannot be applied. %In
Ref.~\cite{Butterworth} considered the reconstruction of top quark,
and eliminated this background by vetoing the events containing a
top quark. Since we have already extracted the ``single jet'' $J$ of
$j_1j_2$ satisfying the conditions (\ref{M_J}) and (\ref{log cut}),
the momentum of the ``single jet'' can be measured. Then we can
combine this ``single jet'' with the remaining jets $j$ (the $b$
jets) to reconstruct the top quark mass. FIG.\,\ref{M_{Jj}} depicts
the invariant mass $M_{Jj}$ distributions for the top quark
background and the $pp\to W^+W^\pm j^f_1j^f_2$ process, which shows
that we can extract the top quark peak by requiring
\cite{Butterworth}
\begin{eqnarray}                                     %(20)
130~{\rm GeV}< M_{Jj} <240~{\rm GeV}. \label{top veto}
\end{eqnarray}
We do it event by event, and veto the events containing the top
quark. This {\it top quark veto} requirement can further suppress
the top quark background. The effect of this veto is listed in the
sixth row in TABLE\,\ref{cut efficiency}.

\subsection{Additional Cuts}

There are two commonly imposed additional cuts to suppress the
backgrounds. The first one is the {\it $p^{}_T$ balance} requirement
\cite{TU-MSU08} (it is called {\it hard $p^{}_T$} in
Ref.~\cite{Butterworth}). Note that the signal process is a {\it
hard process} in which the sum of the transverse momenta ($p^{}_T$)
of the final state particles vanishes ($p^{}_T$ ballance). In the
mentioned QCD backgrounds, there are undetected missing jets which
carry $p^{}_T$, so that summing up the $p^{}_T$ of the detected
final state particles will not vanish. Therefore imposing the
requirement of $p^{}_T$ balance can further suppress this kind of
background. Considering the resolution of $p^{}_T$ measurement
\cite{ATLAS}, we impose the following $p_T$ balance requirement
\cite{TU-MSU08}

\begin{eqnarray}                       %(21)
\sum_i p^{i}_T <\pm15~{\rm GeV}, \label{p_T balance}
\end{eqnarray}
where $p^{i}_T$ is the transverse momentum of the $i$th final state
particle.

Another additional cut commonly used is called {\it minijet veto}.
For the signal process, there is no color exchange between the
forward jet quarks and the $W^\pm$ decay jet $J$. However, color
exchange is expected in the background processes due to the
remnant-remnant interactions, which can produce minijets. Therefore
one can impose the additional cut of {\it minijet veto} by vetoing
the events containing jets other than the signal jet $J$ from
$W^\pm$ decay [satisfying (\ref{p_T(J)}) and (\ref{M_J})] in the
central rapidity region, $|\eta|<2$ \cite{Butterworth}.

The efficiencies of these additional cuts are shown in the last two
rows in TABLE\,\ref{cut efficiency}.

\begin{widetext}

\begin{table}[h]
\label{cut efficiency}
 \caption{ Cut efficiency of the cross sections (in fb) for the
 signal with irreducible background (IB) and other backgrounds with the Higgs boson
mass $m_{H}=115$ GeV, and the anomalous coupling
$f_{W}/\Lambda^2=4.0~ \text{TeV}^{-2}$ (with other anomalous
couplings vanishing) as an example. }\null\vspace{-0.2cm}
 \tabcolsep 12pt
\begin{tabular}{cccccc}
\hline  \hline
 Cuts   & signal with IB &  IB ($f_W=0$)  &WZ+2-jet   & W+3-jet & $t\bar{t}$\\%  & S/B  \\
%        &  (fb)    & (fb)  &  (fb) &   (fb)          &  (fb)     &      \\
\hline\\
   Without cuts  &  210.66 &   338.82  & 1431.67  &  2908923& 407776.84  \\%& 5.90 $\times{10^{-5}}$ \\
Eq.\,(\ref{p_T(l^+)})%(1)$p_{T}(l^{+})>$200GeV
&  34.55  &   36.08   & 36.93  &  9630.86 & 2586.47\\%& 2.27 $\times{10^{-3}}$ \\
Eq.\,(\ref{forward jet cuts})
 &  11.29   &   9.44    & 2.40  &  104.25  & 61.77 \\%2.7  $\times{10^{-2}}$ \\
Eqs.~(\ref{p_T(J)}) and (\ref{M_J})
&  7.01   &   4.12    & 0.12  &  0.10  & 1.09 \\%& 0.54  \\
Eq.\,(\ref{log cut})
&  2.42   &   1.29    & 2.7$\times{10^{-2}}$ &  6.1$\times{10^{-3}}$ & 0.09\\%& 1.19 \\
Eq.\,(\ref{top veto}) and top quark veto
&  2.39   &   1.27    & 2.3$\times{10^{-2}}$ &  4.7$\times{10^{-3}}$ & 0.06\\%& 1.25 \\
Eq.\,(\ref{p_T balance})
&  2.28   &   1.26    &  5$\times{10^{-4}}$   &  2$\times{10^{-4}}$ &  2$\times{10^{-3}}$\\%  & 1.55 \\
Minijet veto     &  2.28   &   1.26   &  -                   &  -                    &  -       \\\\%             & 1.57 \\
%(4)130\text{GeV}<M_{Wj}<240 &  3.21   &   1.96   &  3.74    &  37.68      &  15.95   & 5.38$\times{10^{-2}}$ \\
%(5)Forward jets veto &  1.21   &   0.49   &  0.44    &  0.65       &  0.05    & 7.02$\times{10^{-1}}$ \\
%(6)$M(j_{1}j_{2})$ veto \quad $800\text{GeV}<M(j1j2)<5000\text{GeV}$ &  0.74   &   0.18   &  0.01    &   -         &   -      & 3.06 \\
\hline \hline \label{cut efficiency}
\end{tabular}
\end{table}

\end{widetext}

To illustrate the efficiencies of all these cuts, we list the cross
sections (in fb) for the signal with irreducible background (IB), IB
(obtained from the same process but with $f_W=0$), the QCD
backgrounds, and the top quark background in TABLE\,\ref{cut
efficiency} for $m_H=115$ GeV and $f_W/\Lambda^2=4$ TeV$^{-2}$ (with
other anomalous couplings vanishing) as an example. We see that the
cuts can significantly suppress the backgrounds.  TABLE\,\ref{cut
efficiency} shows that {\it minijet veto} does not affect the
results much because the above cuts have already very efficiently
suppressed the backgrounds. After the cuts, the main
remained background is the irreducible background %(obtained from
%setting $f_W=0$) which is similar to that of the signal but
which is similar to the signal but is not enhanced at high energies
by the momentum dependence of the anomalous couplings.

\section{Numerical Results}

From (\ref{g}) we see that the anomalous couplings $g^{(i)}_{HVV}$
($i=1,2$, $V$ stands for $\gamma,~W^\pm,~Z^0$) are related to four
parameters, namely $f^{}_W,~f^{}_{WW},~f^{}_B,~f^{}_{BB}$. For the
process $pp\to W^+W^\pm j^f_1j^f_2$, except for the small
contributions related to the photon, the main contributions are from
the anomalous couplings of the Higgs boson to the weak gauge bosons,
which is mainly contributed by $f^{}_W$ and $f^{}_{WW}$ since the
contributions from $f^{}_B$ and $f^{}_{BB}$ are suppressed by a
factor of $\sin^2\theta^{}_W$ or $\sin^4\theta^{}_W$ [cf.
Eq.\,(\ref{g})]. In the following, we only take account of the
contributions related to $f^{}_W$ and $f^{}_{WW}$, and neglect the
$f^{}_B$ and $f^{}_{BB}$ contributions (setting
$f^{}_B,f^{}_{BB}=0$). With the above kinematic cuts, We give a full
tree level calculation of the signal and background cross sections,
event numbers, statistical significance [using Eq.\,(\ref{P_B})] for
several values of integrated luminosity with various values of
$f^{}_W/\Lambda^2$ and $f^{}_{WW}/\Lambda^2$ for $m_H$=115, 160, and
200 GeV. In this paper, we only take into account the statistical
uncertainty. The issue related to the systematic error is beyond the
scope of this paper, and we leave it to the experimentalists.

For simplicity, we first make a one-parameter study, i.e.,
considering the cases of $f^{}_W/\Lambda^2$ dominant and
$f^{}_{WW}/\Lambda^2$ dominant separately. We shall discuss the
two-parameter study in the end of this section.

\begin{widetext}

\begin{table}[h]
 \caption{\label{cross-sections} Cross sections (in fb) for
$pp\rightarrow W^{+}W^{\pm}j^f_1j^f_2\rightarrow l^{+}\nu^{}_l
j_1j_2j^f_1j^f_2$ ($l^{+}=e^{+},\mu^{+} $) at the LHC  with various
values of $f^{}_{W}/\Lambda^{2}$ and $\ f^{}_{WW}/\Lambda^{2}$ (in
TeV$^{-2}$) for $m_H=115,~160$ and 200 GeV.}
\begin{ruledtabular}
\begin{tabular}{ccccccccccc}
  $m_{H}$ (GeV)     & & &  &  &$\displaystyle\frac{f^{}_{W}}{\Lambda^2}$ (TeV$^{-2}$) & & &\\
 % \hline
 &-4.0  & -3.0  & -2.0  & -1.0  &  0  &  1.0  & 2.0 & 3.0  & 4.0  \\
\hline
  115  &3.23  &2.91 &1.26 &1.06 &1.19 &1.18 &1.51 &1.82 &2.28
\\
  160  &1.65  &1.32 &1.15 &1.13 &1.22 &1.43 &1.65 &1.77 &2.18
\\
  200  &1.93  &1.86 &1.80 &1.79 &1.82 &2.30 &2.43 &2.53 &2.66
\\
\hline \hline
  $m_{H}$ (GeV)     & & &  &  &$\displaystyle\frac{f^{}_{WW}}{\Lambda^2}$ (TeV$^{-2})$& & &\\
%  \hline
  &-4.0  & -3.0  & -2.0  & -1.0  &  0  &  1.0  & 2.0 & 3.0  & 4.0  \\
\hline
  115 &4.88   &3.11 &1.66 &1.37 &1.19 &1.34 &2.04 &3.34 &5.36\\
  160 &12.35   &4.48 &2.10 &1.36 &1.22 &1.64 &2.70 &4.12 &6.90\\
  200 &11.50   &5.61 &3.27 &2.11 &1.82 &2.26 &2.74 &4.46 &6.94\\
\end{tabular}
\end{ruledtabular}
\end{table}
\end{widetext}

First, we list in TABLE\,\ref{cross-sections} the obtained cross
sections with various values of $f^{}_W/\Lambda^2$ and
$f^{}_{WW}/\Lambda^2$ (in TeV$^{-2}$) for $m_H$=115, 160, and 200
GeV. Note that the positive and negative regions of
$f^{}_W/\Lambda^2$ and $f^{}_{WW}/\Lambda^2$ are not symmetric due
to the interference between the signal and irreducible background
amplitudes. We see that the cross sections are of the order of 1 fb
which are larger than those in the pure leptonic mode [$O(0.1~{\rm
fb})$] \cite{ZKHY03} by and order of magnitude. The largeness of the
cross sections is due to: (i) the branching ratio for $W\to j_1j_2$
is larger than that for $W\to l^+\nu^{}_l$, and (ii) we have
included the process $pp\to W^+W^-j^f_1j^f_2\to
l^+\nu^{}_lj_1j_2j^f_1j^f_2$ as well, and with the improved cuts.

From TABLE\,\ref{cross-sections} we see that for an integrated
luminosity of 100 fb$^{-1}$, there can be of $O(10^2)$ events
detected at the LHC. This not only reduces the statistical
uncertainty relative to that in the pure leptonic mode, but also
provides the possibility of measuring the differential cross
sections. This is the advantage of the semileptonic mode.

Next, we take an integrated luminosity of ${\cal L}_{int}\equiv\int
dt{\cal L}=100$ fb$^{-1}$ to calculated the event numbers and using
%the Poisson probability distribution
the approach of Eq.\,(\ref{P_B}) to find out the sensitivities of
$f^{}_{W}/\Lambda^{2}$ and $\ f^{}_{WW}/\Lambda^{2}$ (in TeV$^{-2}$)
[and the related $g^{(i)}_{HVV}$ (in TeV$^{-1}$) in Eq.\,(\ref{g})]
corresponding to the statistical significance of $1\sigma,~2\sigma$
and $3\sigma$ for $m_H=115,~160$ and 200 GeV. The results are listed
in Eqs.~(\ref{100-115}), (\ref{100-160}), and (\ref{100-200}).

%\begin{widetext}

For $m^{}_H=115$ GeV and ${\cal L}_{int}=100$ fb$^{-1}$
($f^{}_W/\Lambda^2$, $f^{}_{WW}/\Lambda^2$ in TeV$^{-2}$,
$g^{(i)}_{HVV}$ in TeV$^{-1}$), the results are:

\null\vspace{-0.7cm}
\begin{eqnarray}                                %(22)
&&\null\hspace{-0.5cm}1\sigma:\nonumber\\
&&\null\hspace{-0.3cm}-2.0<f^{}_{W}/\Lambda^2<1.2,~~-0.4
<f^{}_{WW}/\Lambda^2<0.8,~~~~\nonumber\\
&&\null\hspace{-0.3cm}-0.053<g^{(1)}_{HWW}<0.032,~-0.042<g^{(2)}_{HWW}<0.021,\nonumber\\
&&\null\hspace{-0.3cm}-0.053<g^{(1)}_{HZZ}<0.032,~~-0.016<g^{(2)}_{HZZ}<0.008,\nonumber\\
&&\null\hspace{-0.3cm}-0.029<g^{(1)}_{HZ\gamma}<0.017,~~-0.018<g^{(2)}_{HZ\gamma}<0.009,\nonumber\\
&&\null\hspace{-0.3cm}-0.005<g^{}_{H\gamma\gamma}<0.002.\nonumber\\
%\end{eqnarray}
%\begin{eqnarray}
&&\null\hspace{-0.5cm}2\sigma:\nonumber\\
&&\null\hspace{-0.3cm}-2.2<f^{}_{W}/\Lambda^2<1.6,~~-1.1
<f^{}_{WW}/\Lambda^2<1.1,\nonumber\\
&&\null\hspace{-0.3cm}-0.058<g^{(1)}_{HWW}<0.042,~-0.058<g^{(2)}_{HWW}<0.058,\nonumber\\
&&\null\hspace{-0.3cm}-0.058<g^{(1)}_{HZZ}<0.042,~~-0.022<g^{(2)}_{HZZ}<0.022,\nonumber\\
&&\null\hspace{-0.3cm}-0.032<g^{(1)}_{HZ\gamma}<0.023,~~-0.024<g^{(2)}_{HZ\gamma}<0.024,\nonumber\\
&&\null\hspace{-0.3cm}-0.007<g^{}_{H\gamma\gamma}<0.007.\nonumber\\
%\end{eqnarray}
%\begin{eqnarray}
&&\null\hspace{-0.5cm}3\sigma:\nonumber\\
&&\null\hspace{-0.3cm}-2.4<f^{}_{W}/\Lambda^2<1.9,~~-1.5
<f^{}_{WW}/\Lambda^2<1.3,\nonumber\\
&&\null\hspace{-0.3cm}-0.063<g^{(1)}_{HWW}<0.050,~-0.068<g^{(2)}_{HWW}<0.079,\nonumber\\
&&\null\hspace{-0.3cm}-0.063<g^{(1)}_{HZZ}<0.050~~-0.026<g^{(2)}_{HZZ}<0.030,\nonumber\\
&&\null\hspace{-0.3cm}-0.035<g^{(1)}_{HZ\gamma}<0.027,~~-0.029<g^{(2)}_{HZ\gamma}<0.033,\nonumber\\
&&\null\hspace{-0.3cm}-0.008<g^{}_{H\gamma\gamma}<0.009.
\label{100-115}
\end{eqnarray}

For $m^{}_H=160$ GeV and ${\cal L}_{int}=100$ fb$^{-1}$
($f^{}_W/\Lambda^2$, $f^{}_{WW}/\Lambda^2$ in TeV$^{-2}$,
$g^{(i)}_{HVV}$ in TeV$^{-1}$), the results are:
\begin{eqnarray}                                %(23)
&&\null\hspace{-0.5cm}1\sigma:\nonumber\\
&&\null\hspace{-0.3cm}-2.7<f^{}_{W}/\Lambda^2<0.3,~~-0.9
<f^{}_{WW}/\Lambda^2<0.2,\nonumber\\
&&\null\hspace{-0.3cm}-0.071<g^{(1)}_{HWW}<0.008,~-0.011<g^{(2)}_{HWW}<0.047,\nonumber\\
&&\null\hspace{-0.3cm}-0.071<g^{(1)}_{HZZ}<0.008,~~-0.004<g^{(2)}_{HZZ}<0.018,\nonumber\\
&&\null\hspace{-0.3cm}-0.039<g^{(1)}_{HZ\gamma}<0.004,~~-0.004<g^{(2)}_{HZ\gamma}<0.020,\nonumber\\
&&\null\hspace{-0.3cm}-0.001<g^{}_{H\gamma\gamma}<0.005.\nonumber\\
%\end{eqnarray}
%\begin{eqnarray}
&&\null\hspace{-0.5cm}2\sigma:\nonumber\\
&&\null\hspace{-0.3cm}-3.4<f^{}_{W}/\Lambda^2<0.9,~~-1.1
<f^{}_{WW}/\Lambda^2<0.5,\nonumber\\
&&\null\hspace{-0.3cm}-0.089<g^{(1)}_{HWW}<0.024,~-0.026<g^{(2)}_{HWW}<0.058,\nonumber\\
&&\null\hspace{-0.3cm}-0.089<g^{(1)}_{HZZ}<0.024,~~-0.010<g^{(2)}_{HZZ}<0.022,\nonumber\\
&&\null\hspace{-0.3cm}-0.049<g^{(1)}_{HZ\gamma}<0.013,~~-0.011<g^{(2)}_{HZ\gamma}<0.024,\nonumber\\
&&\null\hspace{-0.3cm}-0.003<g^{}_{H\gamma\gamma}<0.007.\nonumber\\
%\end{eqnarray}
%\begin{eqnarray}
&&\null\hspace{-0.5cm}3\sigma:\nonumber\\
&&\null\hspace{-0.3cm}-3.8<f^{}_{W}/\Lambda^2<1.5,~~-1.3
<f^{}_{WW}/\Lambda^2<0.8,\nonumber\\
&&\null\hspace{-0.3cm}-0.100<g^{(1)}_{HWW}<0.039,~-0.042<g^{(2)}_{HWW}<0.068,\nonumber\\
&&\null\hspace{-0.3cm}-0.100<g^{(1)}_{HZZ}<0.039~~-0.016<g^{(2)}_{HZZ}<0.026,\nonumber\\
&&\null\hspace{-0.3cm}-0.055<g^{(1)}_{HZ\gamma}<0.022,~~-0.018<g^{(2)}_{HZ\gamma}<0.029,\nonumber\\
&&\null\hspace{-0.3cm}-0.005<g^{}_{H\gamma\gamma}<0.008.
\label{100-160}
\end{eqnarray}

For $m^{}_H=200$ GeV and ${\cal L}_{int}=100$ fb$^{-1}$
($f^{}_W/\Lambda^2$, $f^{}_{WW}/\Lambda^2$ in TeV$^{-2}$,
$g^{(i)}_{HVV}$ in TeV$^{-1}$), the results are:
\begin{eqnarray}                                %(24)
&&\null\hspace{-0.5cm}1\sigma:\nonumber\\
&&\null\hspace{-0.3cm}-3.2<f^{}_{W}/\Lambda^2<0.2,~~-0.7
<f^{}_{WW}/\Lambda^2<0.2,\nonumber\\
&&\null\hspace{-0.3cm}-0.084<g^{(1)}_{HWW}<0.005,~-0.011<g^{(2)}_{HWW}<0.037,\nonumber\\
&&\null\hspace{-0.3cm}-0.084<g^{(1)}_{HZZ}<0.005~~-0.004<g^{(2)}_{HZZ}<0.014,\nonumber\\
&&\null\hspace{-0.3cm}-0.046<g^{(1)}_{HZ\gamma}<0.003,~~-0.004<g^{(2)}_{HZ\gamma}<0.015,\nonumber\\
&&\null\hspace{-0.3cm}-0.001<g^{}_{H\gamma\gamma}<0.004.\nonumber\\
%\end{eqnarray}
%\begin{eqnarray}
&&\null\hspace{-0.5cm}2\sigma:\nonumber\\
&&\null\hspace{-0.3cm}-4.1<f^{}_{W}/\Lambda^2<0.6,~~-1.0
<f^{}_{WW}/\Lambda^2<0.7,\nonumber\\
&&\null\hspace{-0.3cm}-0.108<g^{(1)}_{HWW}<0.016,~-0.037<g^{(2)}_{HWW}<0.053,\nonumber\\
&&\null\hspace{-0.3cm}-0.108<g^{(1)}_{HZZ}<0.016,~~-0.014<g^{(2)}_{HZZ}<0.020,\nonumber\\
&&\null\hspace{-0.3cm}-0.059<g^{(1)}_{HZ\gamma}<0.009,~~-0.015<g^{(2)}_{HZ\gamma}<0.022,\nonumber\\
&&\null\hspace{-0.3cm}-0.004<g^{}_{H\gamma\gamma}<0.006.\nonumber\\
%\end{eqnarray}
%\begin{eqnarray}
&&\null\hspace{-0.5cm}3\sigma:\nonumber\\
&&\null\hspace{-0.3cm}-4.3<f^{}_{W}/\Lambda^2<0.8,~~-1.2
<f^{}_{WW}/\Lambda^2<1.0,\nonumber\\
&&\null\hspace{-0.3cm}-0.113<g^{(1)}_{HWW}<0.021,~-0.053<g^{(2)}_{HWW}<0.063,\nonumber\\
&&\null\hspace{-0.3cm}-0.113<g^{(1)}_{HZZ}<0.021~~-0.020<g^{(2)}_{HZZ}<0.024,\nonumber\\
&&\null\hspace{-0.3cm}-0.062<g^{(1)}_{HZ\gamma}<0.012,~~-0.022<g^{(2)}_{HZ\gamma}<0.027,\nonumber\\
&&\null\hspace{-0.3cm}-0.006<g^{}_{H\gamma\gamma}<0.007.
\label{100-200}
\end{eqnarray}
%\end{widetext}

Eq.\,(\ref{100-115}) is to be compared with the sensitivities in the
pure leptonic mode with $m^{}_H=115$ GeV for an integrated
luminosity of 300 fb $^{-1}$ ($f^{}_W/\Lambda^2$ and
$f^{}_{WW}/\Lambda^2$ are in TeV$^{-2}$) \cite{ZKHY03}.
\begin{eqnarray}                                         %(25)
&&\null\hspace{-0.7cm}1\sigma:\nonumber\\
&&\null\hspace{-0.5cm}-1.0<f^{}_W/\Lambda^2<0.85,~~-1.6<f^{}_{WW}/\Lambda^2,1.6,\nonumber\\
&&\null\hspace{-0.7cm}2\sigma:\nonumber\\
&&\null\hspace{-0.5cm}-1.4<f^{}_W/\Lambda^2<1.2,~~-2.2<f^{}_{WW}/\Lambda^2<2.2,\nonumber\\
&&\null\hspace{-0.7cm}3\sigma:\nonumber\\
&&\null\hspace{-0.5cm}-1.7<f^{}_W/\Lambda^2<1.6,~~-2.9<f^{}_{WW}/\Lambda^2<2.9.
\label{pure leptonic sensitivity}
\end{eqnarray}

Note that $f^{}_W/\Lambda^2$ is more sensitive in the pure leptonic
mode, while $f^{}_{WW}/\Lambda^2$ is more sensitive in the
semileptonic mode. This is because that the process considered in
the pure leptonic mode is only $pp\to W^+W^+j^f_1j^f_2$, while it is
$pp\to W^+W^\pm j^f_1j^f_2$ in the semileptonic mode. Anyway, the
$2\sigma$ sensitivities in the two modes are of the same level.
Since the required integrated luminosity in the pure leptonic mode
is 300 fb$^{-1}$ while it is only 100 fb$^{-1}$ in the semileptonic
mode, the semileptonic mode can reduce the required integrated
luminosity by a factor of 3 relative to the pure leptonic mode. So
the anomalous couplings can be measured to this sensitivity when the
LHC reaches its designed luminosity, 100 fb$^{-1}$/year, or even
earlier. This is quite promising.

\begin{widetext}

\begin{table}[h]
 \caption{\label{50} Numbers of events for $pp\rightarrow W^{+}W^{\pm}j^f_1j^f_2\rightarrow l^{+}\nu^{}_l
j_1j_2j^f_1j^f_2$ ($l^{+}=e^{+},\mu^{+} $) at the LHC for an
integrated luminosity of 50 fb$^{-1}$ with various values of
$f^{}_{W}/\Lambda^{2}$ and $\ f^{}_{WW}/\Lambda^{2}$ (in TeV$^{-2}$)
for $m_H=115,~160$ and 200 GeV. The values of the statistical
significance $\sigma_{stat}$ are shown in the parentheses.}
\begin{ruledtabular}
\begin{tabular}{ccccccccccc}
  $m_{H}$ (GeV)     & & &  &  &$\displaystyle\frac{f^{}_W}{\Lambda^2}$ (TeV$^{-2}$) & & &\\
 % \hline
 &-4.0  & -3.0  & -2.0  & -1.0  &  0  &  1.0  & 2.0 & 3.0  & 4.0  \\
\hline
  115  &162 (13.21)  &146 (11.16) &63 (0.81) &53 (-) &60 (0) &59 (-) &76 (2.12) &91 (4.05) &114 (7.09) \\
  160  &83 (2.75)  &66 (1.09) &58 (-)  &57 (-) &61 (0) &72 (1.58) &83 (2.75) &89 (3.50) &109(6.09)\\
  200  &96 (1.01)  &93 (0.79)&90 (-) &89 (-)&91 (0) &115 (2.54) &121 (3.18) &126 (3.71) &133 (4.39)
\\
\hline \hline
  $m_{H}$ (GeV)     & & &  &  &$\displaystyle\frac{f^{}_{WW}}{\Lambda^2}$ (TeV$^{-2}$) & & &\\
 % \hline
  &-4.0  & -3.0  & -2.0  & -1.0  &  0  &  1.0  & 2.0 & 3.0  & 4.0  \\
\hline
  115 &244 (23.89)   &156 (12.41) &83  (3.06) &69 (1.39)  &60 (0) &67 (1.28)  &102 (5.51) &167 (13.91) &268 (26.99)\\
  160 &618 (71.13)   &224 (20.85) &105 (5.61) &68 (1.18)  &62 (0) &82 (2.75)  &135 (9.42) &206 (18.52) &345 (36.30)\\
  200 &575 (50.14)   &281 (19.56) &164 (7.39) &106 (1.56) &93 (0) &113 (2.17) &137 (4.64) &223 (13.56) &347 (26.44)\\
\end{tabular}
\end{ruledtabular}
\end{table}
\end{widetext}

So far we have concentrated on the study of the detection
sensitivities. In the real world, the actual anomalous coupling(s)
might be larger than the sensitivity bound(s) given above. So
nonvanishing anomalous coupling(s) might even be detected for lower
integrated luminosities at the LHC. Let us take the integrated
luminosity of 50 fb$^{-1}$ as an example. In TABLE\,\ref{50}, we
list the numbers of events for $pp\rightarrow
W^{+}W^{\pm}j^f_1j^f_2\rightarrow l^{+}\nu^{}_l j_1j_2j^f_1j^f_2$ at
the LHC for an integrated luminosity of 50 fb$^{-1}$ with various
values of $f^{}_{W}/\Lambda^{2}$ and $\ f^{}_{WW}/\Lambda^{2}$ (in
TeV$^{-2}$) for $m_H=115,~160$ and 200 GeV. The values of the
statistical significance $\sigma_{stat}$ are shown in the
parentheses.

 Our calculation shows that the sensitivity bounds for $m_H=115\--200$ GeV
 and ${\cal L}_{int}=50$ fb$^{-1}$ are:
 \begin{eqnarray}                     %(26)
 &&1\sigma:\nonumber\\
 &&~~~~-3.5~{\rm TeV}^{-2}\le f^{}_W/\Lambda^2\le 1.3~{\rm
 TeV}^{-2},\nonumber\\
 &&~~~~-0.9~{\rm TeV}^{-2}\le f^{}_{WW}/\Lambda^2\le 0.8~{\rm
 TeV}^{-2},\nonumber\\
 &&3\sigma:\nonumber\\
 &&~~~~-4.5~{\rm TeV}^{-2}\le f^{}_W/\Lambda^2\le 2.4~{\rm TeV}^{-2},\nonumber\\
 &&~~~~
 -2.0~{\rm TeV}^{-2}\le f^{}_{WW}/\Lambda^2\le 1.5~{\rm TeV}^{-2}.
 \label{50-sensitivity}
 \end{eqnarray}

If the anomalous coupling constants in the nature are beyond the
$1\sigma$ bounds in (\ref{50-sensitivity}),
 %$-3.5~{\rm TeV}^{-2}\le f^{}_W/\Lambda^2\le 1.3~{\rm TeV}^{-2}$ or
%$-0.9~{\rm TeV}^{-2}\le f^{}_{WW}/\Lambda^2\le 0.8~{\rm TeV}^{-2}$ ,
the LHC can already detect their effect with several tens to a
hundred of events when the integrated luminosity reaches 50
fb$^{-1}$. This is quite promising since it can be started within
the first couple of years run of the LHC. If they are beyond the
$3\sigma$ bounds,
%$-4.5~{\rm TeV}^{-2}\le f^{}_W/\Lambda^2|\le 2.4$
%TeV$^{-2}$ or $-2.0~{\rm TeV}^{-2}\le f^{}_{WW}/\Lambda^2|\le 1.5$
%TeV$^{-2}$,
the LHC can perform a $3\sigma$ detection for an integrated
luminosity of 50 fb$^{-1}$. If the experiment does not find the
evidence of the anomalous couplings at the LHC for an integrated
luminosity of 50 fb$^{-1}$, it means that $f^{}_W/\Lambda^2$ and
$f^{}_{WW}/\Lambda^2$ are within the $1\sigma$ sensitivity bounds
given in (\ref{50-sensitivity}),
%range $-3.5~{\rm TeV}^{-2}\le
%f^{}_W/\Lambda^2\le 1.3~{\rm TeV}^{-2}$ (or $-0.9~{\rm TeV}^{-2}\le
%f^{}_{WW}/\Lambda^2\le 0.8~{\rm TeV}^{-2}$),
and further detection
with higher integrated luminosity is needed.

Finally we show some results of the two-parameter study.
\begin{figure}[h]                                 %Fig.10
\centering   \bc
\includegraphics[width=9.2cm,height=6cm]{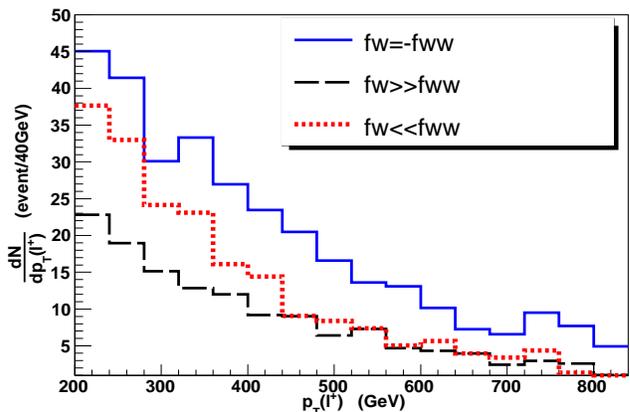}
\ec \vspace{-0.8cm}\caption{\label{lepton p_T distributions} Leptom
transverse momentum distributions in the case of $m^{}_H=115$ GeV
for an integrated luminosity of 100 fb$^{-1}$ taking the cases of
$f^{}_W/\Lambda^2=2~{\rm TeV}^{-2}\gg
f^{}_{WW}/\Lambda^2,~f^{}_{WW}/\Lambda^2=2~{\rm TeV}^{-2}\gg
f^{}_W/\Lambda^2$ and $f^{}_W/\Lambda^2=-f^{}_{WW}/\Lambda^2=2~{\rm
TeV}^{-2}$ as examples. } \label{lepton p_T distributions}
\end{figure}

As mentioned above, with the large cross sections in the
semileptonic mode, we can study differential cross sections which
behave differently for different values of $f^{}_W$ and $f^{}_{WW}$,
so that we can determine $f^{}_W$ and $f^{}_{WW}$ separately from
this information. As an example, we plot, in FIG.\,\ref{lepton p_T
distributions}, the $p^{}_T(l^+)$ distributions for $m^{}_H=115$ GeV and ${\cal L}_{int}$=100 fb$^{-1}$
contributed by three
different sets of $f^{}_W/\Lambda^2$ and $f^{}_{WW}/\Lambda^2$ from
different regions in the two-parameter space, namely the cases of
$f^{}_W/\Lambda^2=2~{\rm TeV}^{-2}\gg
f^{}_{WW}/\Lambda^2,~f^{}_{WW}/\Lambda^2=2~{\rm TeV}^{-2}\gg
f^{}_W/\Lambda^2$, and $f^{}_W/\Lambda^2=-f^{}_{WW}/\Lambda^2=2~{\rm
TeV}^{-2}$.
%and $f^{}_W/\Lambda^2=f^{}_{WW}/\Lambda^2=2~{\rm TeV}^{-2}$,
%
%In FIG.~\ref{lepton p_T distributions} we plot the $p^{}_T$
%distributions of the charged lepton for three example cases of
%$f^{}_W$ and $f^{}_{WW}$, say $f^{}_W/\Lambda^2=2~{\rm TeV}^{-2}\gg
%f^{}_{WW}/\Lambda^2,~f^{}_{WW}/\Lambda^2=2~{\rm TeV}^{-2}\gg
%f^{}_W/\Lambda^2$, and $f^{}_W/\Lambda^2=-f^{}_{WW}/\Lambda^2=2~{\rm
%TeV}^{-2}$
%and $f^{}_W/\Lambda^2=f^{}_{WW}/\Lambda^2=2~{\rm TeV}^{-2}$,
% for $m^{}_H=115$ GeV and ${\cal L}_{int}$=100 fb$^{-1}$.
We see that the three $p^{}_T(l^+)$
distributions are different and quite distinguishable especially in
the region near 200 GeV. Since the cross section is more sensitive
to $f^{}_{WW}/\Lambda^2$ than to $f^{}_W/\Lambda^2$, the curve of
the $f^{}_{WW}/\Lambda^2=2~{\rm TeV}^{-2}\gg f^{}_W/\Lambda^2$ case
lies significantly higher than that of the
$f^{}_{W}/\Lambda^2=2~{\rm TeV}^{-2}\gg f^{}_{WW}/\Lambda^2$ case.
%They are easy to distinguish.
From Eq.\,(\ref{g}) we see that
$f^{}_W/\Lambda^2$ appears in the formulae always with a positive
sign, while $f^{}_{WW}/\Lambda^2$ appears always with a negative
sign. So that in the case of
$f^{}_W/\Lambda^2=-f^{}_{WW}/\Lambda^2=2~{\rm TeV}^{-2}$, these two
contributions are constructive, and thus this curve lies well above
the two former curves.
%In the case of
%$f^{}_W/\Lambda^2=f^{}_{WW}/\Lambda^2=2~{\rm TeV}^{-2}$, the
%contributions of $f^{}_W/\Lambda^2$ and $f^{}_{WW}/\Lambda^2$ are
%destructive, so that the curve of this case lies below all the other
%three curves. Since the two contributions only cancel each other
%partly, this curve lies not much lower than that of the
%$f^{}_{W}/\Lambda^2=2~{\rm TeV}^{-2}\gg f^{}_{WW}/\Lambda^2$ case.
%However, it is still distinguishable from the other three curves.
Therefore measuring both the cross section and the the $p^{}_T(l^+)$
distribution may help to separately determine the two parameters
$f^{}_W$ and $f^{}_{WW}$ to a certain precision. If there is no
characteristic signal for new physics model found before this
measurement, the values of $f^{}_W$ and $f^{}_{WW}$ may serve as a
clue for probing the underlying theory of new physics. This is an
advantage of the semileptonic mode over the pure leptonic mode.

\section{Conclusion}

In this paper, we have given a full tree level study of the test of
anomalous gauge couplings [cf. Eqs.\,(\ref{l:eff}) and (\ref{g})] at
the LHC via the $WW$ scattering processes $pp\to W^+W^+j^f_1j^f_2$
and $pp\to W^+W^-j^f_1j^f_2$ in the semileptonic mode $W^+\to
l^+\nu^{}_l,~W^+~(W^-)\to j_1j_2$. Through out this paper, we take
into account only the statistical uncertainty. The issue of
systematic error is beyond the scope of this paper, and we leave it
to the experimentalists.

Both signals and backgrounds are calculated at the hadron level with
suitably imposed kinematic cuts to suppress the backgrounds. As we
mentioned in Sec.\,III-A, the signal and irreducible background
should to calculated together to guarantee gauge invariance. The
efficiencies of the cuts are shown in TABLE\,\ref{cut efficiency}
which shows that the cuts (\ref{p_T(l^+)})$\--$(\ref{p_T balance})
can suppress the QCD backgrounds and the $t\bar t$ background quite
efficiently. After the cuts, the main background remained is the
irreducible background.

The obtained cross sections for $m^{}_H=115,~160$ and 200 GeV in the
ranges $|f^{}_W/\Lambda^2|\le 4$ and $|f^{}_{WW}/\Lambda^2|\le 4$
are listed in TABLE\,\ref{cross-sections}. Because of the largeness
of the branching ratio $B(W^\pm\to j_1j_2)$, the contributions of
both $pp\to W^+W^+j^f_1j^f_2$ and $pp\to W^+W^-j^f_1j^f_2$, and the
improved cuts, the cross sections are as large as of $O({\rm
1\,fb})\--O(10\,{\rm fb})$. So that for an integrated luminosity of
100 fb$^{-1}$, hundreds of events can be detected at the LHC.

As mentioned in Sec. IV that the $pp\to W^+W^\pm j^f_1j^f_2$
processes are mainly sensitive to two anomalous coupling constants,
$f^{}_W/\Lambda^2$ and $f^{}_{WW}/\Lambda^2$. We first made a
one-parameter study, i.e., considering the cases of
$f^{}_W/\Lambda^2$ dominant and $f^{}_{WW}/\Lambda^2$ dominant
separately. Taking the integrated luminosity of 100 fb$^{-1}$ as an
example, the obtained results of the sensitivity ranges of
$f^{}_W/\Lambda^2$, $f^{}_{WW}/\Lambda^2$ and the corresponding
$g^{(i)}_{HVV}$'s for $1\sigma,~2\sigma$ and $3\sigma$ detections
are listed in Eqs.\,(\ref{100-115}) to (\ref{100-200}) for
$m^{}_H=115$ GeV, 160 GeV and 200 GeV. These are of the same level
as those in the pure leptonic mode for an integrated luminosity of
300 fb$^{-1}$. Thus for the same level of sensitivity, the
semileptonic mode can reduce the required integrated luminosity by a
factor of 3.

If the actual anomalous coupling constants in nature are not so
small, it can even be measured with a low luminosity as 50
fb$^{-1}$. The obtained event numbers and statistical significance
$\sigma_{stat}$ for an luminosity of 50 fb$^{-1}$ are listed in
TABLE\,\ref{50} which shows that a detection with around $O(100)$
events can be performed at the LHC for an integrated luminosity of
50 fb$^{-1}$ if the anomalous coupling constants in the nature are
larger than the $1\sigma$ bounds given in
Eq.\,(\ref{50-sensitivity}). This can be done within the first couple
of years run of the LHC. So it is quite promising. If the detected
result is consistent with the SM value at the LHC for an integrated
luminosity of 50 fb$^{-1}$, it means that $f^{}_W/\Lambda^2$ and
$f^{}_{WW}/\Lambda^2$ are within the $1\sigma$ sensitivity bounds
(\ref{50-sensitivity}), and further detection with higher integrated
luminosity is needed.

We have also made a simple two-parameter study considering
$f^{}_W/\Lambda^2$ and $f^{}_{WW}/\Lambda^2$ simultaneously. With
the hundreds of events for ${\cal L}_{int}=100$ fb$^{-1}$, it is
possible to measure the $p^{}_T$ distribution of the charged lepton
experimentally. We plotted in FIG.\,\ref{lepton p_T distributions}
the $p^{}_T(l^+)$ distributions for $m^{}_H=115$ GeV and ${\cal
L}_{int}=100$ fb$^{-1}$ corresponding to $f^{}_W/\Lambda^2=2\,{\rm
TeV}^{-2}\gg f^{}_{WW}/\Lambda^2,~f^{}_{WW}/\Lambda^2=2\,{\rm
TeV}^{-2}\gg f^{}_W/\Lambda^2$, and
$f^{}_W/\Lambda^2=-f^{}_{WW}/\Lambda^2=2\,{\rm TeV}^{-2}$
% and
%$f^{}_W/\Lambda^2=f^{}_{WW}/\Lambda^2=2~{\rm TeV}^{-2}$
as examples. It shows that the three distributions are quite
distinguishable. Therefore measuring both the total cross section
and the $p^{}_T(l^+)$ distribution may determine the two parameters
$f^{}_W/\Lambda^2$ and $f^{}_{WW}/\Lambda^2$ separately to certain
precision. This may provide a clue for figuring out the underlying
theory of new physics beyond the SM if no other characteristic
signal of the new physics is found before that measurement.
%The $1\sigma\--5\sigma$ bounds for
%$m^{}_H=115$ GeV and ${\cal L}_{int}=100$ fb$^{-1}$ in the
%$f^{}_W/\Lambda^2$-$f^{}_{WW}/\Lambda^2$ plane when
%$f^{}_W/\Lambda^2$ and $f^{}_{WW}/\Lambda^2$ are positive is shown
%in FIG.~\ref{contours}.

In summary, the process $pp\to W^+W^\pm j^f_1j^f_2\to
l^+\nu^{}_lj^{}_1j^{}_2j^f_1j^f_2$ at the LHC can provide a
sensitive test of the anomalous gauge couplings of the Higgs boson
showing the effect of new physics beyond the SM. The experiment can
start the test for an integrated luminosity around 50 fb$^{-1}$, and
can measure the total cross section and the $p^{}_T$ distributions
of the charged lepton to certain precision for an integrated
luminosity of 100 fb$^{-1}$. With such measurements, it is possible
to determine the two main parameters $f^{}_W/\Lambda^2$ and
$f^{}_{WW}/\Lambda^2$ of the anomalous couplings separately, which
may provide a clue for figuring out the underlying theory of new
physics.

\null\vspace{0.2cm} \centerline{\bf ACKNOWLEDGMENTS}

%\null\vspace{0.1cm}
 We would like to thank Yuanning Gao for useful
discussions. This work is supported by National Natural Science
Foundation of China under Grant Nos. 10635030, 10875064, 10705017
and 10435040.

%\newpage
\bibliography{000}

\end{document}